\documentclass[12pt]{article}

\usepackage{amssymb}
\usepackage{amsmath}

\usepackage{graphicx}

\begin{document} %


\title{Probing electromagnetic moments of the tau lepton in PbPb collisions at the FCC-hh}

\author{
S.C. \.{I}nan\thanks{Electronic address: sceminan@cumhuriyet.tr}
\\
{\small Department of Physics, Sivas Cumhuriyet University, 58140,
Sivas, Turkey}
\\
{\small and}
\\
A.V. Kisselev\thanks{Electronic address:
alexandre.kisselev@ihep.ru} \\
{\small Division of Theoretical Physics, A.A. Logunov Institute for
High Energy Physics,}
\\
{\small NRC ``Kurchatov Institute'', 142281, Protvino, Russia}}

\date{}

\maketitle

\begin{abstract}
A production of a pair of tau leptons in ultra-peripheral PbPb
collisions at the FCC-hh collider is examined. The 95\% C.L.
exclusion limits, as well as 3$\sigma$ and 5$\sigma$ sensitivity
limits on the anomaly magnetic moment of the tau lepton $a_\tau$ and
its electric dipole moment $d_\tau$ are obtained. A comparison with
bounds on $a_\tau$ and $d_\tau$ for other future colliders are
given.
\end{abstract}

\maketitle


\section{Introduction} %

The evidence of a new lepton $\tau$ was found half a century ago
\cite{Perl:1975} and confirmed two years later. A study of $\tau$
anomalous magnetic moment represents an excellent possibility to
access new physics beyond the Standard model (SM)
\cite{Perl:1998,Pich:2014}. A magnetic moment $\vec{\mu}$ in an
externally produced magnetic field $\vec{B}$ has a potential energy
\begin{equation}\label{U_in_B}
U = - \vec{\mu}\vec{B} \;.
\end{equation}
In the Born approximation the magnetic moment is given by
\begin{equation}\label{magnetic_moment}
\vec{\mu} = g \!\left( \frac{e\hbar}{2mc}\right) \!\vec{s} \;,
\end{equation}
where $e$, $m$, and $\vec{s}$ is an electric charge, mass, and spin
of the particle, and $g$ being a gyromagnetic factor.  The Dirac
equation predicts $g=2$. An anomalous magnetic moment,
\begin{equation}\label{a}
a = \frac{g-2}{2} \;,
\end{equation}
describes a deviation of $g$ from the Dirac value.

The first QED correction to it was calculated for the electron, $a_e
= \alpha/(2\pi)= 0.00116141$, where $\alpha = e^2/(4\pi)$ is the
fine-structure constant \cite{Schwinger:1948}. The $a_e$ is measured
with a precision of 0.13 ppt \cite{a_e:2023}, and the muon anomalous
magnetic moment $a_\mu$ is measured by the ``Muon g-2''
collaboration with a precision of 127 ppb \cite{a_mu:2021}. This
first-order correction to $a_l$ is flavor and mass independent. That
is why in the one-loop approximation $a_\tau = \alpha/(2\pi)$. The
SM contribution to the tau lepton anomalous magnet moment $a_\tau$
from higher loop corrections is equal to
\cite{Passera:2007,Eidelman:2007}
\begin{equation}\label{a_tau_QED}
a_\tau^{\mathrm{SM}} =  117721(5) \times 10^{-8} \;.
\end{equation}
It is a sum of QED, electroweak and hadronic contributions. The
higher corrections to a lepton $a_l$ were calculated in
\cite{Passera:2007,Eidelman:2007}.

The short proper  $\tau$ lifetime does not allow to apply the spin
procession method used for the $\mu$ case.%
\footnote{However, see \cite{Fomin:2019,Fu:2019} and references
therein.}
That is why it is necessary to probe the sensitivity of the
$\tau$-pair differential and total cross sections to the
electromagnetic moments of the tau lepton. The experimental 95\%
C.L. constraints on $a_\tau$ obtained from the $Z \rightarrow
\tau\tau\gamma$ decay \cite{L3:1998,OPAL:1998} and the process
$e^+e^- \rightarrow e^+e^-\tau^+\tau^-$ at the LEP look like
\cite{DELPHI:2004}
\begin{align}\label{a_tau_LEP}
-0.052 &< a_\tau < 0.058 \ (\mathrm{L3}) \;,
\nonumber \\
-0.068 &< a_\tau < 0.065 \ (\mathrm{OPAL}) \;,
\nonumber \\
-0.052 &< a_\tau < 0.013 \ (\mathrm{DELPHI}) \;.
\end{align}

The ATLAS PbPb 95\% confidence-level interval for $a_\tau$
\cite{ATLAS:2023,ATLAS:2026} is equal to
\begin{equation}\label{a_tau_ATLAS_Pb}
-0.057 < a_\tau < 0.024 \;.
\end{equation}
The recent CMS limits on $a_\tau$ from PbPb collisions
\cite{CMS:2026} are
\begin{equation}\label{a_tau_CMS_Pb}
-0.030 < a_\tau < 0.032 \;.
\end{equation}
The 95\% C.L. constraints on $a_\tau$ obtained from the pp
measurements by the ATLAS look like \cite{ATLAS:2025}
\begin{equation}\label{a_tau_ATLAS_p}
-0.0024 < a_\tau < 0.0047 \;,
\end{equation}
and the CMS $pp$ result at 95\% C.L. is the following
\cite{CMS:2024}
\begin{equation}\label{a_tau_CMS_p}
-0.0022 < a_\tau < 0.0041 \;,
\end{equation}
see also \cite{Volkotrub:2024}-\cite{{Clawson:2025}}. Thus, the pp
collisions provide a large complementary dataset to the PbPb
measurements.

Prospects for measuring $a_\tau$ by the ALICE Collaboration in Run
3, which benefits from high statistics and improved systematics
uncertainties, was discussed in \cite{ALICE:2025}.

Deviations from $a_\tau = 0$ arise from the radiative corrections
and can also stem from effects of new physics. In many BSM scenarios
where New Physics (NP) couples to the lepton masses, the NP
contributions are expected to scale as $a_\tau \sim
m_l^2/\Lambda_{\mathrm{NP}}^2$. As a result, NP contributions are
expected to scale as $a_\tau/a_\mu = m_\tau^2/m_\mu^2 = 286$. Thus,
the larger $\tau$ mass makes the hadronic contributions
significantly larger than the case of the muon.

Another valuable contribution to interaction of the photon with the
tau lepton is CP violation \cite{Christenson:1964} generated by
$\tau$ electric dipole moment. An electric dipole moment of the
lepton $\vec{d}$ in an external electric field $\vec{E}$ acquires a
potential energy
\begin{equation}\label{U_in_E}
U = - \vec{d}\vec{E} \;,
\end{equation}
with $\vec{d} = d \vec{s}$. In a non-relativistic limit, it follows
from the Dirac equation with an additional Pauli term that is
responsible for the electric dipole moment,
\begin{equation}\label{Dirac_eq}
\left( i\hbar c\hat{\partial} - e\hat{A} - m)\Psi - \frac{i}{2} d
\gamma^5 \sigma^{\mu\nu}F_{\mu\nu} \right)\psi = 0 \;,
\end{equation}
where $\sigma^{\mu\nu} = (i/2)(\gamma^\mu\gamma^\nu -
\gamma^\nu\gamma^\mu)$.

The upper 95\% CL limits on the $\tau$ electric dipole moment
$d_\tau$ obtained by the LEP collaborations are the following
\cite{L3:1998}-\cite{DELPHI:2004}
\begin{align}\label{d_tau_LEP}
&|d_\tau| < 3.1\times 10^{-16} \mathrm{\ e\ cm} \ (\mathrm{L3}) \;,
\nonumber \\
&|d_\tau| < 3.7\times 10^{-16} \mathrm{\ e\ cm} \ (\mathrm{OPAL})
\;,
\nonumber \\
&|d_\tau| < 3.7\times 10^{-16} \mathrm{\ e\ cm} \ (\mathrm{DELPHI})
\;.
\end{align}
The values of $d_\tau$ set by the ARGUS collaboration are
\cite{ARGUS:2000}
\begin{align}\label{d_tau_ARGUS}
& \mathfrak{Re}(d_\tau) = (1.6 \pm 1.9)\times 10^{-16} \mathrm{\ e\
cm} \
\nonumber\\
& \mathfrak{Im}(d_\tau) = (-0.2 \pm 0.8)\times 10^{-16} \mathrm{\ e\
cm} \;,
\end{align}
which translate into the following 95\% CL upper limits
\cite{ARGUS:2000}
\begin{align}\label{d_tau_ARGUS_bounds}
& |\mathfrak{Re}(d_\tau)| < 4.6 \times 10^{-16} \mathrm{\ e\ cm} \;,
\nonumber\\
& |\mathfrak{Im}(d_\tau)| < 1.8\times 10^{-16} \mathrm{\ e\ cm} \;.
\end{align}
The results obtained by the Belle collaboration are stronger
\cite{Belle:2003,Belle:2022},
\begin{align}\label{d_tau_Belle}
& \mathfrak{Re}(d_\tau) = (-0.62 \pm 0.63)\times 10^{-17} \mathrm{\
e\ cm} \;,
\nonumber\\
& \mathfrak{Im}(d_\tau) = (-0.40 \pm 0.32)\times 10^{-17} \mathrm{\
e\ cm} \;.
\end{align}
The 95\% confidence intervals becomes \cite{Belle:2022}
\begin{align}\label{d_tau_Belle_bounds}
-1.85 \times 10^{-17} < \mathfrak{Re}(d_\tau) < 0.61 \times 10^{-17}
\mathrm{\ e\ cm} \;,
\nonumber\\
-1.03 \times 10^{-17} < \mathfrak{Re}(d_\tau) < 0.23 \times 10^{-17}
\mathrm{\ e\ cm} \;.
\end{align}
The CMS 95\% CL limit on $d_\tau$ obtained from the pp measurements
is equal to \cite{CMS:2024}
\begin{equation}\label{d_tau_CMS_p}
|d_\tau| < 2.9 \times 10^{-17} \mathrm{\ e\ cm} \;.
\end{equation}
Finally, the 95\% CL bound obtained by the ATLAS collaboration for
the PbPb collisions at $\sqrt{s} = 5.02$ TeV is equal to
\cite{ATLAS:2026}
\begin{equation}\label{d_tau_ATLAS_Pb}
|d_\tau| < 2.7 \times 10^{-16} \mathrm{\ e\ cm} \;.
\end{equation}
These results are the first constraints obtained using Pb+Pb
collisions. As in the case of $a_\tau$, the LHC constraint on
$d_\tau$ from the PbPb collisions is an order of magnitude weaker
than that from the $pp$ collisions. In what follows, it is assumed
that $d_\tau$ is given in units of [e cm].

Large $a_\tau$ and $d_\tau$ can be realized in models with SUSY
\cite{Ellis:1982}-\cite{Ilakovac:2014}, leptoquarks
\cite{Barr:1986,Ma:1992}, vector like multiplets
\cite{Ibrahim:2010}, tau compositeness
\cite{Silverman:1983,Goertz:2022}, in extended technicolor model
\cite{Appelquist:2004}, two Higgs doublet model \cite{Iltan:2001} or
Kaluza-Klein theories \cite{Calmet:2002}-\cite{Kadosh:2010}.

Measurements of the anomalous magnetic and electric dipole moments
of the tau lepton can be made using $pp$ collisions
\cite{Huang:1998}-\cite{Haisch:2024} or ultra-peripheral collisions
(UPCs) at the LHC \cite{ATLAS:2023,CMS:2026},
\cite{Aguila:1991}-\cite{Lu:2026}. In particular, NLO QED and
electroweak corrections to the cross section were estimated
\cite{Shao:2025,Dittmaier:2025}. Probing $\tau$ electromagnetic
properties at future lepton colliders was studied in
\cite{Peressutti:2012}, \cite{Bernreuther:1993}-\cite{Huang:2026}.
In \cite{Koksal:2019,Gutierrez-Rodriguez:2022} a search for $a_\tau$
and $d_\tau$ at $eh$ colliders was presented, and constraints on
$a_\tau$ and $d_\tau$ via tau pair production at the muon colliders
were obtained in \cite{{Koksal:2019_2}}-\cite{Denizli:2025}.

For more details on an actual state of the tau magnetic and electric
dipole moments, see \cite{Shokr:TAU2025}-\cite{Gogniat:TAU2025}.

\section{$\gamma\gamma\rightarrow\tau^+\tau^-$ scattering at the FCC-hh} %

Because $a_\tau$ is poorly limited experimentally, there is a
striking room for BSM physics. As mentioned in Introduction, the
very short lifetime of the tau lepton ($2.9 \times 10^{-13}$ s)
precludes the use of the precession frequency measurement method as
done in the $\mu$ case. In the present paper, the method used is to
probe the sensitivity to $a_\tau$ of a cross section of the
$\gamma\gamma\rightarrow\tau^+\tau^-$ scattering in lead-lead
collisions at the FCC-hh \cite{FCC:V_3,FCC:V_1}.

The collision under consideration,
\begin{equation}\label{process}
PbPb \rightarrow Pb \,\gamma\gamma \, Pb \rightarrow Pb
\,\tau^+\tau^- \, Pb \;,
\end{equation}
benefits from the $Z^4$ photon flux enhancement, with $Z = 82$, and
possibility to impose exclusivity requirements thanks to the lack of
pileup events compared to $pp$ collisions. It compensates the lower
integrated luminosity compared with that available in $pp$
collisions, while the requirement of an exclusive final state with
only tau-decay products allows a better control of the background
processes than in a case of proton-proton collisions.

In the equivalent photon approximation (EPA) \cite{Budnev:1975} the
differential cross section of \eqref{process} can be factorized as
\begin{equation}\label{cs}
d\sigma = \int\limits_{\tau_{\min}}^{\tau_{\max}}
\!\!\frac{d\tau}{\tau}
\!\!\int\limits_{\omega_{\min}}^{\omega_{\max}}
\!\!\frac{d\omega}{\omega} f_{\gamma/N}(\omega)
f_{\gamma/N}(\tau/\omega) \,d\hat{\sigma} (\gamma\gamma\rightarrow
\tau^+\tau^-) \;,
\end{equation}
where $\omega_{\max} = E_N - m_N, \tau_{\max} = (E_N - m_N)^2$,
$\omega$ is the photon energies emitted from the nucleus, $E_N$ is
the energy of the ion beams, and $m_N$ is the nucleus mass. The
quantity $4\tau$ coincides with the center-of-mass energy squared of
the process $\gamma\gamma\rightarrow\tau^+\tau^-$. As for the lower
limits on variables $\omega$ and $\tau$ in \eqref{cs}, they are
given by $\omega_{\min} =  \tau/\omega_{\max}, \tau_{\min} = p_t^2 +
m^2$, where $p_t$ is the transverse momenta of the outgoing leptons,
and $m$ is their mass.

In the relativistic limit the equivalent spectrum of the photon from
the nucleus $N$ with the charge $Z$ and atomic number $A$ is given
by \cite{Jackson:QED}-\cite{Baltz:2008}
\begin{equation}\label{dist_gamma_nucleus}
f_{\gamma/N}(\omega) = \frac{2Z^2\alpha}{\pi} \!\left[ \xi K_0(\xi)
K_1(\xi) - \frac{\xi^2}{2} ( K_1^2(\xi) - K_0^2(\xi) ) \right] ,
\end{equation}
where $\xi = \omega/E_R$, $E_R = E_N/(m_N R_A)$, and $R_A =
1.2A^{1/3}$ fm is the radius of the nucleus. $K_0(x)$ ($K_1(x)$) is
the modified Bessel function of the second kind of order zero (one).
For the lead-lead collisions at the FCC-hh ($Z = 82$, $A = 208$,
$\sqrt{s} = 100$ TeV), the energy beam is equal to $E_N = 4097.6$
TeV, and the energy per nucleon is $\sqrt{s_{NN}} = \sqrt{s}\,(Z/A)
= 39.4$ TeV.

The tau lepton-photon coupling can be parameterized by three form
factors, $F_1(q^2)$, $F_2(q^2)$, and $F_3(q^2)$
\cite{Grimus:1988,Bernreuther:1997},
\begin{equation}\label{tau_vertex}
\langle \tau(p')|J^{\mathrm{em}}_\mu(0)|\tau(p)\rangle = -
\bar{u}(p')\Gamma_\mu u(p) \;,
\end{equation}
\begin{equation}\label{photon-tau_vertex}
\Gamma_\mu = -ie \!\left[ F_1(q^2)\gamma_\mu +
F_2(q^2)\frac{i}{2m}\sigma_{\mu\nu} q^\nu +
F_3(q^2)\frac{1}{2m}\sigma_{\mu\nu} q^\nu\gamma_5 \right] ,
\end{equation}
where $q_\nu$ is the momentum transfer of the photon, $m$ is the
mass of the tau lepton. $F_1(q^2)$ is known as the charge (Dirac)
form factor whereas $F_2(q^2)$ as the magnetic (Pauli) form factor.
Both are P- and C-even.

The form factor $F_3(q^2)$ is related to the electric dipole moment.
A non-zero electric dipole moment violates both parity (P) and
time-reversal (T) symmetries and, assuming CPT invariance, is
therefore CP-violating. Since the SM predicts extremely small lepton
electric dipole moments, any experimentally observable non-zero
value would constitute compelling evidence for new sources of BSM CP
violation.

In QED all the $F_i(0)$ ($i=1,2,3$) are real thanks to hermiticity
of the electromagnetic current. However, in the framework of the SM
the form factors $F_i(0)$ have imaginary parts. In the limit $q^2
\rightarrow 0$ the form factors describe the static properties of
the tau lepton,
\begin{equation}\label{zero_q2_MM}
F_1(0) = 1, \quad F_2(0) = a_\tau, \quad F_3(0) =
-\frac{2m}{e}d_\tau \;.
\end{equation}
It corresponds to an interaction Lagrangian
\begin{equation}\label{L_eff}
\mathcal{L}_{\mathrm{int}} = \bar{\tau}\!\left[ e\gamma_\mu A^\mu +
\frac{\sigma_{\mu\nu}}{2} \left( a_\tau \frac{e}{2m} - id_\tau
\gamma_5 \right)\!F^{\mu\nu} \right]\!\tau \;.
\end{equation}
For $q^2\neq 0$ the form factors $F_i(q^2)$ have infrared
divergences due to ordinary QED radiative corrections. The
electromagnetic moments \eqref{photon-tau_vertex} at nonzero $q^2$
can be examined in the framework of effective Lagrangian approach,
see Appendix~A.

The amplitude of the photon-induced process
$\gamma\gamma\rightarrow\tau^+\tau^-$ is given by a sum of t- and
u-channel diagrams. Correspondingly, the amplitude squared (after
averaging over initial photon states and summation over final states
of tau leptons) looks like
\begin{equation}\label{M2}
|M|^2 = |M_t|^2 + |M_u|^2 + |M_t^* M_u + M_t M_u^*|^2 \;,
\end{equation}
where
\begin{align}\label{Mt2}
|M_t|^2 &= \frac{8\pi \alpha^2}{m^4(t - m^2)^2}
        \Big[
        48 F_{1}^{3}F_{2}(m^{2}-t)(m^{2}+s-t)m^{4}
        \nonumber\\
        &\quad -16F_{1}^{4}(3m^{4}-sm^{2}+t(s+t))m^{4}
        \nonumber\\
        &\quad +2F_{1}^{2}(m^{2}-t) \big(
        F_{2}^{2}\!\big(17m^{4}+(22s-26t)m^{2}+t(9t-4s)\big)
        \nonumber\\
        &\quad +F_{3}^{2}\!\big(17m^{2}+4s-9t\big)(m^{2}-t)\big) m^{2}
        \nonumber\\
        &\quad +12F_{1}F_{2}(F_{2}^{2}+F_{3}^{2})s(m^{3}-mt)^{2}
        \nonumber\\
        &\quad -(F_{2}^{2}+F_{3}^{2})^{2}(m^{2}-t)^{3}(m^{2}-s-t)
        \Big] ,
\end{align}

\begin{align}\label{Mu2}
|M_u|^2 &= -\frac{8\pi \alpha^2}{m^4(u - m^2)^2} \Big[
        48 F_{1}^{3}F_{2}(m^{4}+(s-2t)m^{2}+t(s+t))m^{4}
        \nonumber\\
        &\quad +16F_{1}^{4}(7m^{4}-(3s+4t)m^{2}+t(s+t))m^{4}
        \nonumber\\
        &\quad
        +2F_{1}^{2}(m^{2}-t)\!\big(
        F_{2}^{2}\!\big(m^{4}+(17s-10t)m^{2}+9t(s+t)\big)
        \nonumber\\
        &\quad +F_{3}^{2}(m^{2}-9t)(m^{2}-t-s)\big)m^{2}
        \nonumber\\
        &\quad
        +(F_{2}^{2}+F_{3}^{2})^{2}(m^{2}-t)^{3}(m^{2}-s-t)
        \Big] ,
\end{align}

\begin{align}\label{Mtu2}
|M_t^* M_u + M_t M_u^*|^2  &= \frac{16\pi \alpha^2}{m^2(t - m^2)(u -
m^2)}
 \Big[
        -16F_{1}^{4}(4m^{6}-m^{4}s)
        \nonumber\\
        &\quad + 8F_{1}^{3}F_{2}m^{2}(6m^{4}-6m^{2}(s+2t)-s^{2}+6t^{2}+6st)
        \nonumber\\
        &\quad + F_{1}^{2}\!\Big(
        F_{2}^{2}\big(16m^{6}-m^{4}(15s+32t)
        \nonumber\\
        &\quad + m^{2}(-15s^{2}+14st+16t^{2})+st(s+t)\big)
        \nonumber\\
        &\quad +F_{3}^{2}\big(16m^{6}-m^{4}(15s+32t)
        \nonumber\\
        &\quad + m^{2}(-5s^{2}+14st+16t^{2})+st(s+t)\big)
        \Big) \nonumber\\
        &\quad -4F_{1}F_{2}(F_{2}^{2}+F_{3}^{2})s(m^{4}+ m^{2}(s-2t)+t(s+t))
        \nonumber\\
        &\quad -4F_{1}F_{3}(F_{2}^{2}+F_{3}^{2})(2m^{2}-s-2t)\,
        \epsilon_{\mu\nu\rho\sigma}k_{1}^{\mu}k_{2}^{\nu}p_{1}^{\rho}p_{2}^{\sigma}
        \nonumber\\
        &\quad -2(F_{2}^{2}+F_{3}^{2})^{2}s(m^{4}-2tm^{2}+t(s+t))
        \Big] .
\end{align}
Here $k_1, k_2$ are the incoming photon momenta, $p_1, p_2$ are
outgoing tau lepton moments. $s, t, u$ are Mandelstam variables, $s
=(k_1 + k_2)^2$, $t=(k_1 - p_1)^2$, $u=(k_1 - p_2)^2$, and $s+t+u =
2m^2$. As one can see, the interference term \eqref{Mtu2} contains
CP-odd Levi-Civita tensor multiplied by odd powers of $F_3$.
However, its contribution to the cross section is zero for the
CP-even initial two-photon state \cite{Atag:2010}. Therefore, we
expect that magnitudes of negative and positive parts of limits on
$d_\tau$ will be the same. Our expressions \eqref{Mt2}-\eqref{Mtu2}
coincides with those obtained previously in
\cite{Atag:2010,Koksal:2018_1,Koksal:2019}.%
\footnote{Note that in refs.~\cite{Koksal:2018_1,Koksal:2019} the
term with the Levi-Civita symbol is absent in eq.~\eqref{Mtu2}.}

From explicit dependence of amplitudes squared
\eqref{Mt2}-\eqref{Mtu2} on $F_2$ and $F_3$, one can expect that the
bounds on $a_\tau$ will be asymmetric, while the bounds on $d_\tau$
will be symmetric. However, after strong cuts are applied, both
bounds become symmetric (see
eqs.~\eqref{SS_escl_limits}-\eqref{SS_disc_5sigma} below).

In the process $PbPb \rightarrow Pb \,(\gamma\gamma\rightarrow
\tau^+\tau^-) \, Pb$ the virtuality of the emitted photons are
$|q^2| \sim 1/R_A^2$, where $R_A = 1.2A^{1/3} \mathrm{\ fm} =
6.09A^{1/3}$ GeV$^{-1}$ is the radius of the nucleus. For the lead
nucleus ($A=208$), it gives $|q^2| \sim 8 \cdot 10^{-4}$ GeV$^2$.
Thus, the photons emitted in this elastic process are almost
on-shell, and the produced tau leptons are back-to-back in the
azimuthal direction. That is why we can replace the form factors in
eqs.~\eqref{Mt2}-\eqref{Mtu2} by their values at $q^2 = 0$
\eqref{zero_q2_MM}. Other approaches to extract $a_\tau$ and
$d_\tau$, not using $\gamma\gamma\rightarrow\tau\tau$ scattering,
have been proposed for DY production, $e^+e^-$ collisions, and fixed
target experiments with bent crystals \cite{Fomin:2019, Fu:2019}.
But in them the photon virtualities are far from zero, and the form
factors $F_i(q^2)$ are suppressed.

We assume that one of the outgoing tau leptons (in what follows,
denoted as $\tau_l$) decays as $\tau^\pm \rightarrow l^\pm + \nu_l +
\nu_\tau$, with $l = e, \mu$ (leptonic $\tau$ decay $\tau_l$), while
the other tau lepton decays as $\tau^\pm \rightarrow \pi^\pm + \pi^0
+ \nu_\tau$ (hadronic $\tau$ decay $\tau_h$). The branching ratios
of these decays are the following \cite{PDG}
\begin{align}\label{tau_brahcn}
\mathrm{Br}(\tau^\pm \rightarrow e^\pm + \nu_e + \nu_\tau) &= 17.8\%
\;,
\nonumber \\
\mathrm{Br}(\tau^\pm \rightarrow \mu^\pm  + \nu_\mu + \nu_\tau) &=
17.4\% \;,
\nonumber \\
\mathrm{Br}(\tau^\pm \rightarrow \pi^\pm + \pi^0 + \nu_\tau) &=
25.5\% \;.
\end{align}
Thus, the number of signal events is calculated via $\sigma(PbPb
\rightarrow Pb\,\tau^+\tau^-Pb) \times \mathrm{B}(\tau^\pm
\rightarrow l^\pm + \nu_l + \nu_\tau) \times \mathrm{B}(\tau^\pm
\rightarrow \pi^\pm + \pi^0 + \nu_\tau) \times
\mathcal{L}_{\mathrm{int}}\times \varepsilon_\tau \times A$, where
$\sigma(PbPb \rightarrow Pb\,\tau^+ \tau^- Pb)$ is the total cross
section of the process \eqref{process}, $\mathcal{L}_{\mathrm{int}}$
is the integrated luminosity, $\varepsilon_\tau$ is the
tau-triggering efficiency of the FCC-hh detector, and $A$ is the
acceptance. We take $\mathcal{L}_{\mathrm{int}} = 110$
nb$^{-1}$~\cite{Yang:2024,Jiang:2024}, and $\varepsilon_\tau = 0.95$
($p_t > 100$ GeV) \cite{tau_eff:2026} (see Fig.~10). The acceptance
is taken into account by the rapidity cut (see below).

In Fig.~\ref{fig:atcsptcut} a dependence of the cross section on
minimal transverse momenta of the outgoing tau leptons $p_{t,\min}$
is presented. The SM cross-section drops sharply with increasing
$p_{t,\min}$. Since the tracker acceptances of the FCC-hh detectors
are expected to be $|\eta| < 4.0$ \cite{FCC:V_1}, we impose the same
cut on the outgoing tau leptons rapidity, $|\eta| < 4.0$. Here the
``SM cross section'' means the cross section calculated with $a_\tau
= d_\tau = 0$.

The background processes are $\gamma\gamma\rightarrow e^+e^-$,
$\gamma\gamma\rightarrow \mu^+\mu^-$, $\gamma\gamma\rightarrow
\mathrm{\ q\bar{q}}$ ($q = u,d,s,c$), and diffractive events. Among
them the $\gamma\gamma\rightarrow \mu^+\mu^-$ process where one of
the $\mu$ radiates a photon is the major source of background. The
probability of the misidentification of an electron as $\tau_h$ is
equal to $P(e\rightarrow \tau_h) = 8.8\times10^{-5}$
($8.3\times10^{-4}$) for $p_t < 100$ GeV ($p_t > 100$ GeV)
\cite{CMS:JINST_2025}, where $p_t$ is the transverse momentum of the
tau lepton. The probability of a muon misidentification was defined
to be $P(\mu\rightarrow \tau_h) = 2.8\times10^{-4}$
($1.1\times10^{-3}$) for $p_t < 100$ GeV ($p_t > 100$ GeV)
\cite{CMS:JINST_2025}.

The contributions from the SM processes
$\gamma\gamma\rightarrow\tau^+\tau^-$ and $\gamma\gamma\rightarrow
\mu^+\mu^-(\gamma)$ were checked by the ATLAS collaboration
\cite{ATLAS:2026}. The observed data yields have demonstrated the
low background level of the background $\gamma\gamma\rightarrow
\mu^+\mu^-(\gamma)$ (16\%) in comparison to the
$\gamma\gamma\rightarrow\tau\tau$ process taken with $a_\tau =
d_\tau = 0$. On the other hand, as it follows from Figs.~1-2, at
$p_t \gtrsim 150$ GeV the SM di-tau production is at least one order
of magnitude lower than the $\gamma\gamma\rightarrow\tau^+\tau^-$
signal with non-zero $a_\tau$ and/or $d_\tau$. It indicates that the
$\gamma\gamma\rightarrow\mu^+\mu^-(\gamma)$ background can be
neglected. As a result, an account for the $\gamma\gamma\rightarrow
\mu^+\mu^-(\gamma)$ processes will not change our bounds on the
magnetic moments of the tau lepton, if we assume that the
experimental misidentification performances of the FCC-hh detectors
will be at least not worse than that of the ATLAS/CMS detectors at
the LHC.

As for the processes $\gamma\gamma\rightarrow \mathrm{\ q\bar{q}}$,
they have a significantly larger charged-particle multiplicity than
the signal and hence this background is fully reducible by
exclusivity requirements \cite{Beresford:2020,Dyndal:2020}. The
contribution of the diffractive photonuclear events is about 2\% of
the $\tau^+\tau^-$ sample as shown in \cite{ATLAS:2023}.

\begin{figure}[htb]
\begin{center}
\includegraphics[scale=0.5]{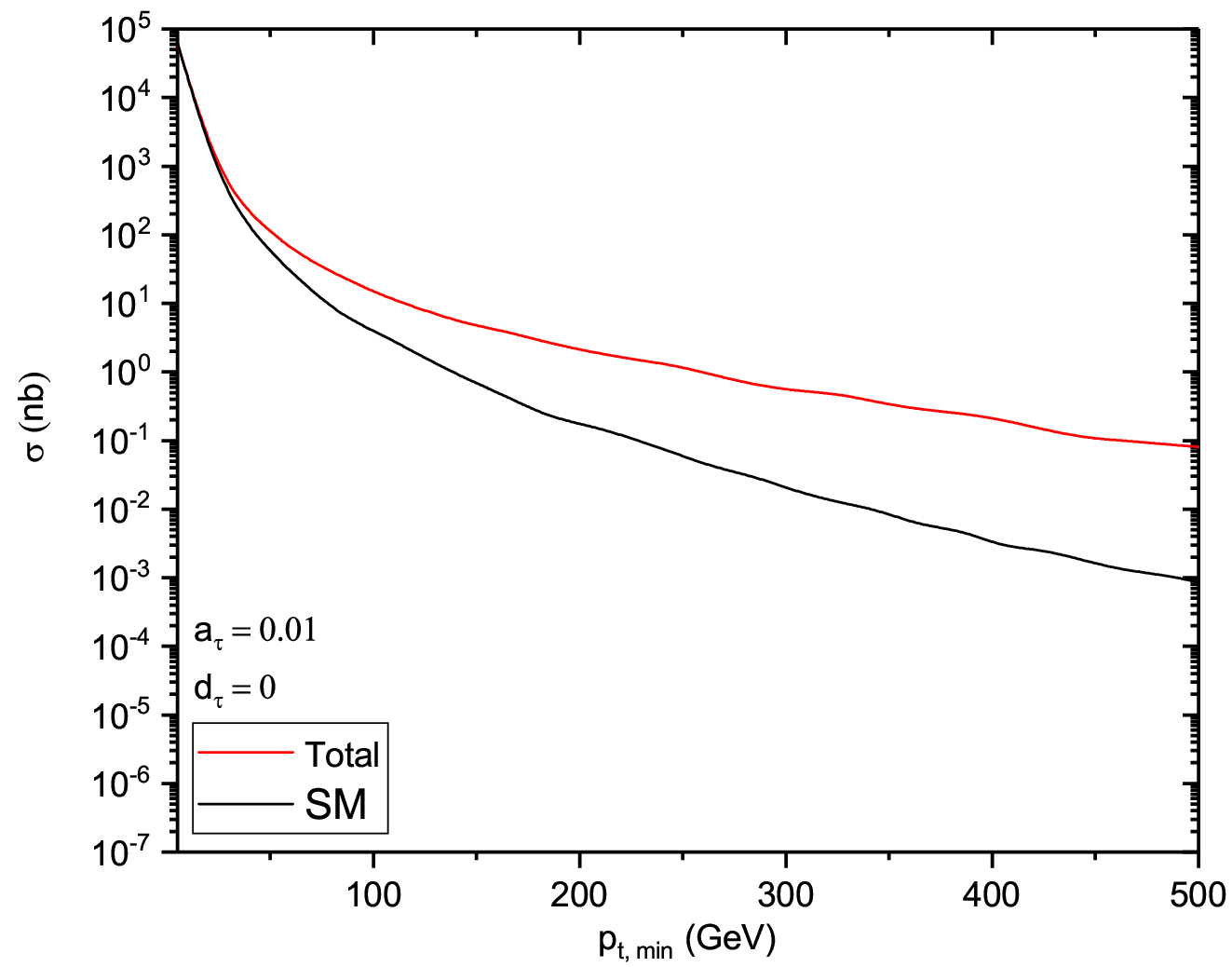}
\caption{The differential cross section of the collision $PbPb
\rightarrow Pb \,(\gamma\gamma\rightarrow \tau^+\tau^-) \, Pb$ at
the FCC-hh versus minimal transverse momenta of the outgoing tau
leptons $p_{t,\min}$. The tau lepton electromagnetic moments are
taken to be $a_\tau = 0.01$, $d_\tau = 0$.}
\label{fig:atcsptcut}
\end{center}
\end{figure}
%
\begin{figure}[htb]
\begin{center}
\includegraphics[scale=0.5]{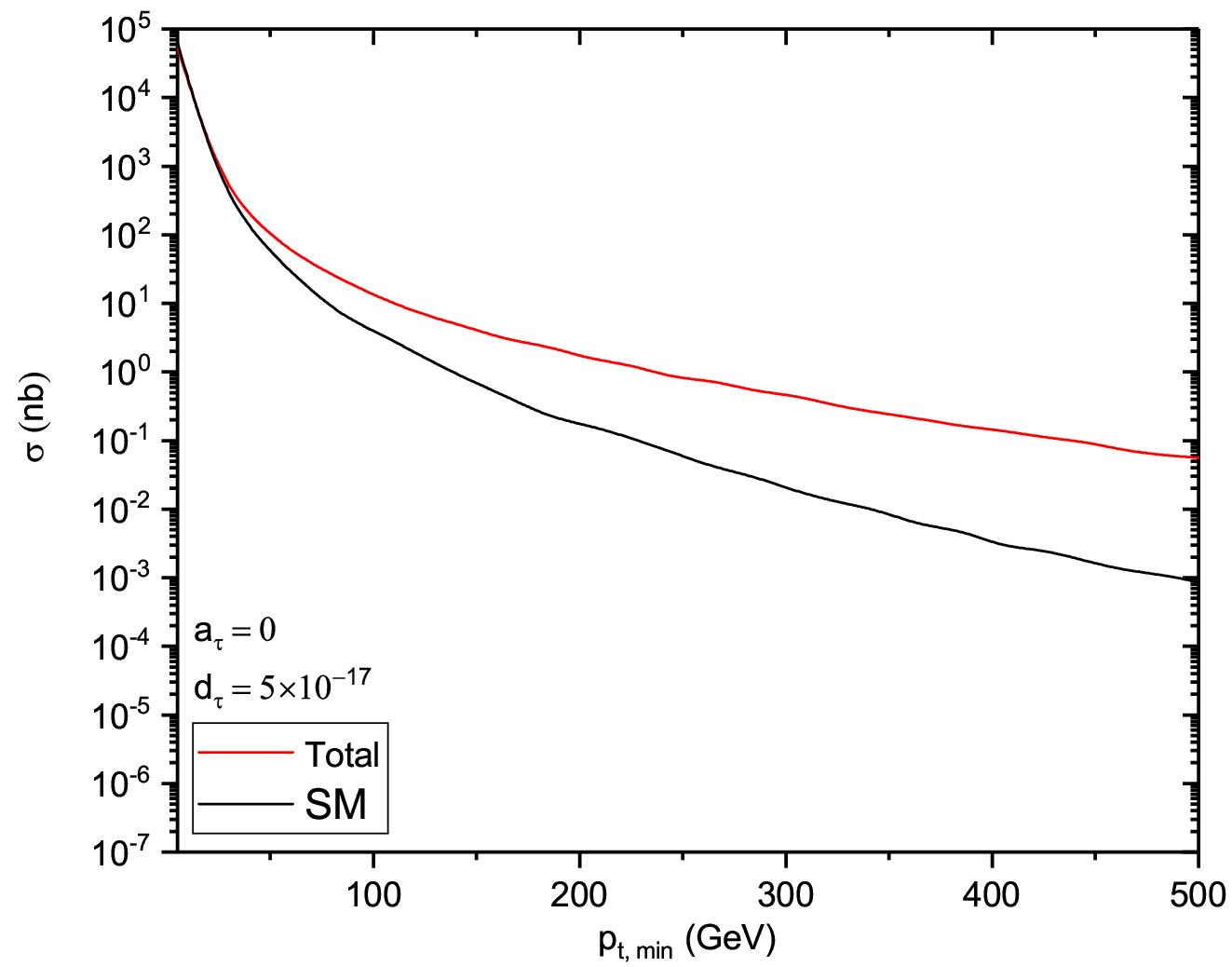}
\caption{The same as in Fig.~~\ref{fig:atcsptcut}, but for the case
$a_\tau = 0$, $d_\tau = 5\times10^{-17}$.}
\label{fig:dtcdptcut}
\end{center}
\end{figure}

Let $S$ and $B$ be the total number of signal and SM events, and
$\delta$ is the percentage systematic uncertainty. Then the
exclusion significance looks like
\cite{Cowan:2011}-\cite{Zhang:2020}
\begin{align}\label{S_excl}
SS_{\mathrm{excl}} = \bigg\{ &2 \left[ S - B \ln\!\left( \frac{B + S
+ x}{2B} \right) - \frac{1}{\delta^2}\ln\!\left( \frac{B - S +
x}{2B}
\right) \right]
\nonumber \\
&- (B + S - x) \left( 1 + \frac{1}{\delta^2B} \right) \bigg\}^{1/2}
,
\end{align}
where
\begin{equation}\label{x}
x = \sqrt{ (S + B)^2 - \frac{4S\delta^2B^2}{(1 + \delta^2 B)} } \;.
\end{equation}
In the limit $\delta \rightarrow 0$ we get from \eqref{S_excl}
\begin{equation}\label{SS_excl_zero_delta}
SS_{\mathrm{excl}} = \sqrt{2[S - B \ln(1 + S/B)]} \;.
\end{equation}
We define the regions $SS_{\mathrm{excl}} \geq 1.645$ as regions
that can be excluded at the 95\% C.L.

\begin{figure}[htb]
\begin{center}
\includegraphics[scale=0.5]{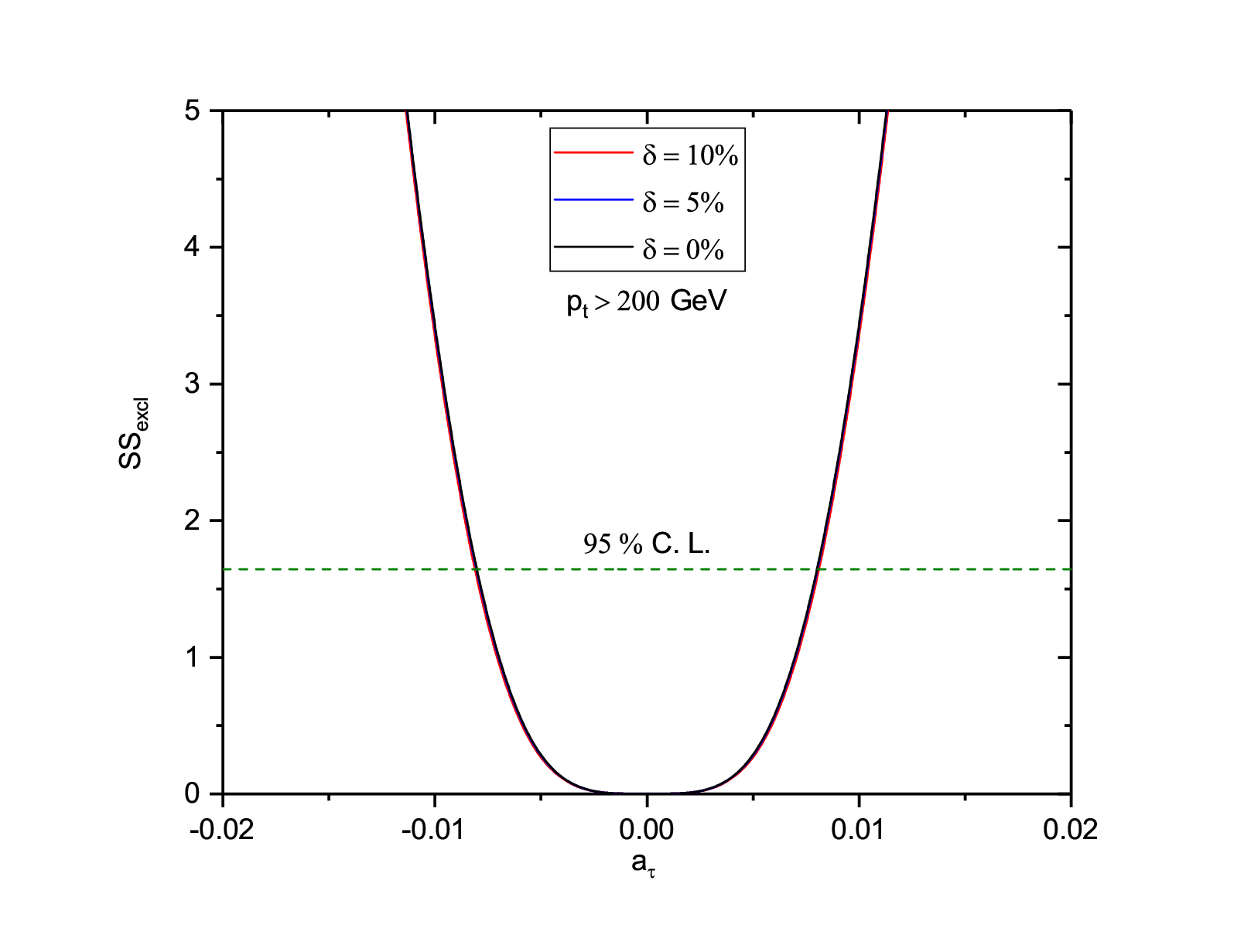}
\caption{The exclusion significance $SS_{\mathrm{excl}}$ as a
function of the anomalous magnetic moment of the tau lepton with the
systematic uncertainty $\delta$ in PbPb collision at the FCC-hh. The
$\tau$ electric dipole moment is fixed and equal to $d_\tau = 0$.}
\label{fig:SSexcl_at_pt200_rpt6_dall}
\end{center}
\end{figure}

\begin{figure}[htb]
\begin{center}
\includegraphics[scale=0.5]{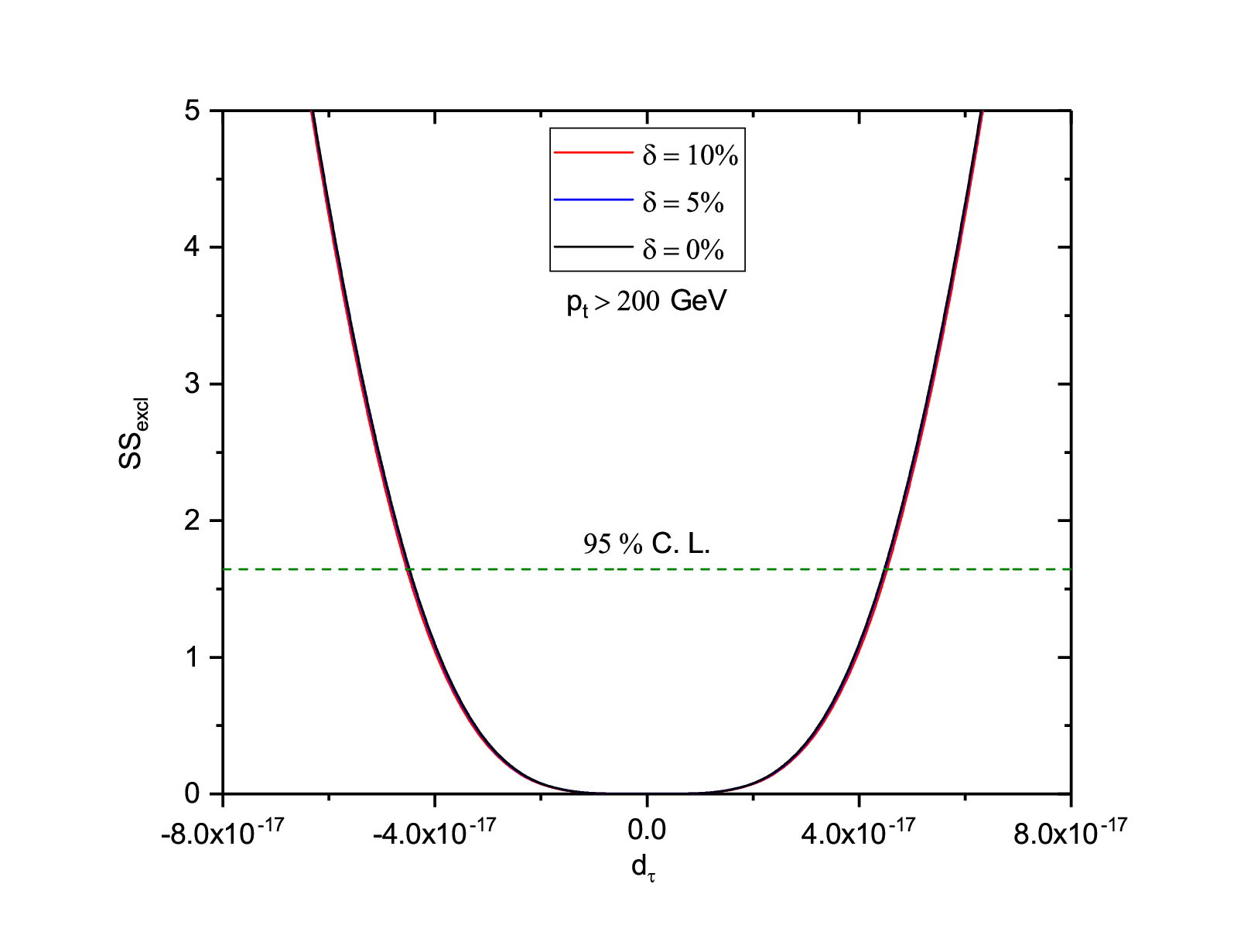}
\caption{The exclusion significance $SS_{\mathrm{excl}}$ as a
function of the electric dipole moment of the tau lepton with the
systematic uncertainty $\delta$ in PbPb collision at the FCC-hh. The
$\tau$ anomalous magnetic moment is fixed and equal to $a_\tau =
0$.} \label{fig:SSexcl_dt_pt200_rpt6_dall}
\end{center}
\end{figure}

As it follows from eqs.~\eqref{M2}-\eqref{Mtu2}, the total cross
sections and, consequently, the number of events $S$ are sensitive
to the electromagnetic form factors of the $\tau$ lepton (anomalous
magnetic moment $a_\tau$, and electric dipole moment $d_\tau$). As a
result, the significance should depend on values of $a_\tau$ and
$d_\tau$. It is demonstrated by
Figs.~\ref{fig:SSexcl_at_pt200_rpt6_dall}-\ref{fig:SSexcl_dt_pt200_rpt6_dall}
where the exclusion significance is presented as a function of the
anomalous magnetic moment and electric dipole moment of the tau
lepton, respectively.

We have explicitly evaluated an impact of systematic uncertainties
at the level of $\delta = 5\%$ and $10\%$ as well. For the applied
selection $p_t > 200$ GeV, the SM background becomes negligibly
small, and the resulting exclusion significance practically
coincides with those obtained for $\delta = 0\%$. It results from
the relation
\begin{equation}\label{S_excl_exp}
SS_{\mathrm{excl}} \simeq \sqrt{2S} \left(1 - \frac{B^2
\delta^2}{4S} + \frac{B^3 \delta^4}{6S} \right)
\end{equation}
valid if $B \ll S$ and $B\delta^2 < 1$. For this reason, in
Figs.~\ref{fig:SSexcl_at_pt200_rpt6_dall} and
\ref{fig:SSexcl_dt_pt200_rpt6_dall} the curves corresponding to
different values of $\delta$ practically coincide.

%

\begin{figure}[htb]
\begin{center}
\includegraphics[scale=0.5]{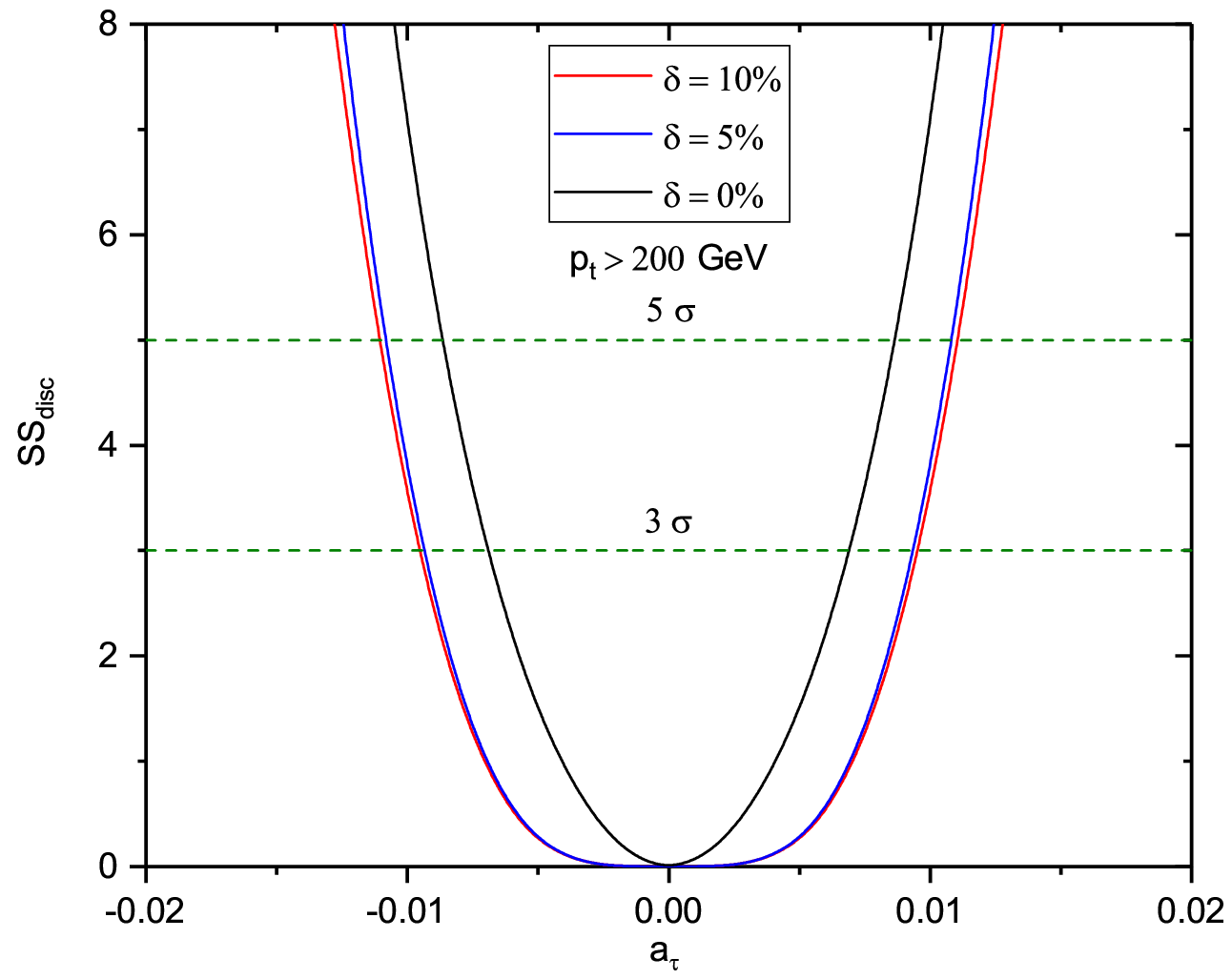}
\caption{The discovery significance $SS_{\mathrm{disc}}$ as a
function of the anomalous magnetic moment of the tau lepton with the
systematic uncertainty $\delta$ in PbPb collision at the FCC-hh. The
$\tau$ electric dipole moment is fixed and equal to $d_\tau = 0$.}
\label{fig:SSdisc_at_pt200_rpt6_dall}
\end{center}
\end{figure}

\begin{figure}[htb]
\begin{center}
\includegraphics[scale=0.5]{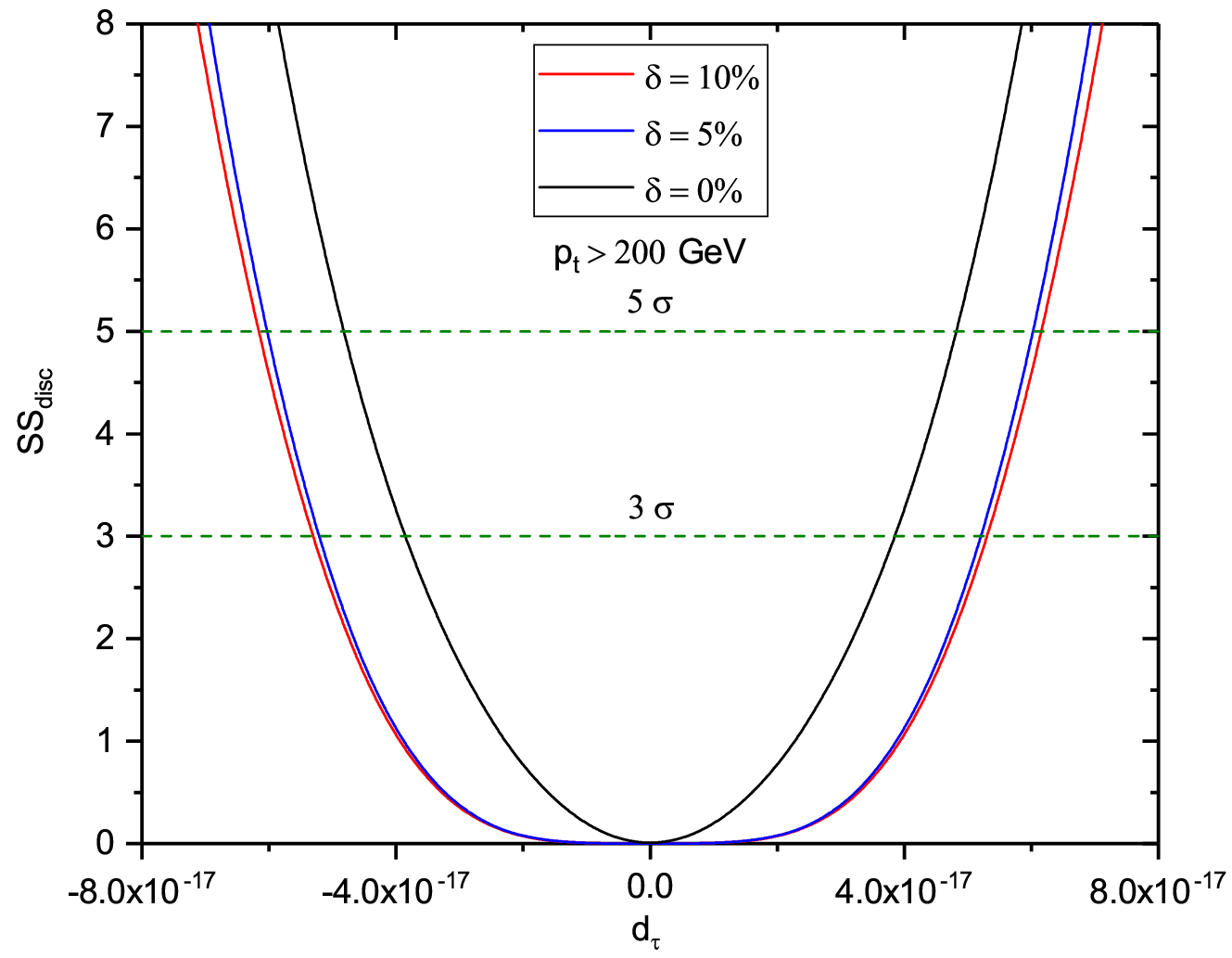}
\caption{The discovery significance $SS_{\mathrm{disc}}$ as a
function of the electric dipole moment of the tau lepton with the
systematic uncertainty $\delta$ in PbPb collision at the FCC-hh. The
$\tau$ anomalous magnetic moment is fixed and equal to $a_\tau =
0$.} \label{fig:SSdisc_dt_pt200_rpt6_dall}
\end{center}
\end{figure}

The discovery significance is defined as
\cite{Cowan:2011}-\cite{Zhang:2020}
\begin{align}\label{S_disc}
SS_{\mathrm{disc}} = \bigg\{ &2\left[ (S + B) \ln\!\left( \frac{(B +
S)(1 + \delta^2 B)}{B + \delta^2 B(S + B)} \right) \right]
\nonumber
\\
&- \frac{1}{\delta^2}\ln\!\left( 1 + \frac{\delta^2 S}{1 + \delta^2
B} \right) \bigg\}^{1/2} \;.
\end{align}
In particular, in the limit $\delta \rightarrow 0$ we find that
\begin{equation}\label{SS_disc_zero_delta}
SS_{\mathrm{disc}} = \sqrt{2[(S + B) \ln(1 + S/B) - S]} \;.
\end{equation}
We classify the region with $SS_{\mathrm{disc}} > 3(5)$ as
discoverable region at $3\sigma(5\sigma)$. The discovery
significance depending on the anomalous magnetic moment and electric
dipole moment of the tau lepton can be found in
Figs.~\ref{fig:SSdisc_at_pt200_rpt6_dall} and
\ref{fig:SSdisc_dt_pt200_rpt6_dall}, respectively. The curves
correspond to different value of the statistical uncertainty
$\delta$.
%

From Figs.~\ref{fig:SSexcl_at_pt200_rpt6_dall},
\ref{fig:SSexcl_dt_pt200_rpt6_dall} we find the 95\% C.L. exclusion
bounds,
\begin{equation}\label{SS_escl_limits}
|a_\tau| \leq 0.0084 \;, \quad |d_\tau| \leq 4.48\times 10^{-17}
\mathrm{\ e\ cm} \;.
\end{equation}
Correspondingly, from Fig.~\ref{fig:SSdisc_at_pt200_rpt6_dall},
\ref{fig:SSdisc_dt_pt200_rpt6_dall} we get both 3$\sigma$
sensitivity limits,
\begin{equation}\label{SS_disc_3sigma}
|a_\tau| \leq 0.0068 \;, \quad |d_\tau| \leq 3.86\times 10^{-17}
\mathrm{\ e\ cm} \;,
\end{equation}
as well as 5$\sigma$ sensitivity limits,
\begin{equation}\label{SS_disc_5sigma}
|a_\tau| \leq 0.0087 \;, \quad |d_\tau| \leq 4.83\times 10^{-17}
\mathrm{\ e\ cm} \;.
\end{equation}

Although the PbPb center-of-mass energy increases significantly from
the LHC to the FCC-hh, the collider sensitivity to the
electromagnetic magnetic moments does not scale proportionally. This
is mainly because the equivalent-photon luminosity is limited by the
finite nuclear radius $R_A$ and increases only moderately with a
beam energy. In addition, to suppress the SM background, we have
imposed the stringent selection cut on transverse momenta of the
outgoing tau leptons ($p_t > 200$ GeV) which also substantially reduces
the signal yield. Consequently, the statistical improvement remains
modest, leading to an expected enhancement of the limits on $a_\tau$
by only a factor of 3 relative to the LHC PbPb projections
\eqref{a_tau_ATLAS_Pb}, \eqref{a_tau_CMS_Pb}. Our upper bound on
$|d_\tau|$ is stronger than the LHC bound \eqref{d_tau_ATLAS_Pb} by
a factor of 6.

In our calculations, we have used the EPA photon flux
\eqref{dist_gamma_nucleus} based on the standard electric dipole
form factor (EDFF). The second photon flux is the charge form factor
(ChFF) which contains an explicit dependence on the photon
transverse momentum \cite{Vidovic:1993,Shao:2022,Shao:2025}. At the
LHC, the $\gamma\gamma \rightarrow \tau^+ \tau^-$ NLO cross sections
computed with the ChFF flux are larger than those obtained with the
EDFF flux by about 23\%, for the ultraperipheral PbPb collisions at
$\sqrt{s} = 5.02$ TeV \cite{Shao:2025}. Then as it follows from
eqs.~\eqref{SS_excl_zero_delta} and \eqref{SS_disc_zero_delta}, our
predictions will change by about 11\%.

\section{Conclusions} %

In this study, we have examined the sensitivity of PbPb UPC at the
FCC-hh to the anomalous electromagnetic moments of the tau lepton
through the process $PbPb \rightarrow Pb \,\gamma\gamma \, Pb
\rightarrow Pb \,\tau^+\tau^- \, Pb$. This work takes advantage of
the large enhancement of the photon flux proportional to $Z^4$ and
the clean experimental environment provided by exclusive final
states in heavy ion collisions. Since in UPCs photon virtualities
$|q^2| < 10^{-3}$ GeV$^2$, we obtain constraints on the anomalous
form factors of the tau lepton in a static limit.

Our results \eqref{SS_escl_limits}-\eqref{SS_disc_5sigma} can be
compared with bounds on the electromagnetic moments of the tau
lepton at other future colliders. In \cite{Gutierrez-Rodriguez:2022}
the bounds for the $e^- p \rightarrow e^-\tau\bar{\tau}\gamma p$
process at the FCC-he with the energy of 10 TeV and integrated
luminosity of $L = 1000$ fb$^{-1}$ at the 95\% C.L. have been
obtained
\begin{equation}\label{prediction_FCC-he}
a_\tau = [-0.00265, 0.00246] \;, \quad |d_\tau| = 1.437\times
10^{-17} \mathrm{e \ cm} \;.
\end{equation}
The expected 95\% C.L. limits for the CLIC operating in the mode
$e^+e^- \rightarrow e^+\tau\bar{\tau}\gamma e^-$ at the energy of 3
TeV and integrated luminosity of $L = 3000$ fb$^{-1}$ look like
\cite{Gutierrez-Rodriguez:2022}
\begin{equation}\label{prediction_CLIC}
a_\tau = [-0.00128, 0.00105] \;, \quad |d_\tau(\mathrm{e \ cm})| =
0.64394\times 10^{-17} \mathrm{\ e\ cm} \;.
\end{equation}
The 95\% C.L. bounds on $a_\tau$ and $d_\tau$ were also estimated
for the $\mu^+\mu^- \rightarrow \mu^+\tau\bar{\tau}\mu^-$ collision
at a future 6 TeV muon collider with $L = 710$ fb$^{-1}$
\cite{Koksal:2019_2}
\begin{equation}\label{prediction_muon_collider_1}
a_\tau = [-0.00242, 0.00071] \;, \quad |d_\tau| = 0.731\times
10^{-17} \;.
\end{equation}
For 10 TeV muon collider and $L = 10$ ab$^{-1}$ the obtained bounds
are stronger \cite{Denizli:2025},
\begin{equation}\label{prediction_muon_collider_2}
a_\tau = [-0.000263, 0.000265] \;, \quad |d_\tau| = 0.147\times
10^{-17}  \mathrm{\ e\ cm} \;.
\end{equation}

Although the projected sensitivities are weaker than those expected at
future lepton colliders, the FCC-hh in PbPb mode gives a
theoretically robust, complementary and independent probe of tau
electromagnetic moments. In particular, heavy ion UPCs offer a
unique combination of large photon luminosities, suppressed hadronic
backgrounds, and distinct systematic uncertainties.  An importance
of heavy-ion collisions, which act as a ``photon-photon collider''
of extreme intensity, was recently emphasized in
\cite{Vignaroli:2026}.

The results obtained here show that the FCC-hh operating in
heavy-ion mode can significantly expand the physical reach of tau
lepton precision studies. Together with other collider studies, PbPb
UPCs constitute a crucial component of examining physics beyond the
SM through tau electromagnetic moments.



\setcounter{equation}{0}
\renewcommand{\theequation}{A.\arabic{equation}}

\setcounter{section}{0}
\renewcommand{\thesection}{A.\arabic{section}}

\section*{Appendix A. Constraints on magnetic and electric dipole moments in SMEFT} %

For $q^2 \neq 0$ the form factors $F_2(q^2)$ and $F_3(q^2)$ in
\eqref{photon-tau_vertex} are not actually the anomalous
electromagnetic moments $a_\tau$ and $d_\tau$. Indeed, it is not
possible that the colliding photons are exactly on-shell. In such a
case, an effective Lagrangian approach can be applied. In the SM
effective field theory (SMEFT) \cite{Buchmuller:1986,Brivio:2019}
the magnetic and electric dipole moments are modified by two
dimension-6 operators,
\begin{equation}\label{L_SMEFT}
\mathcal{L} = \frac{C_{\tau B}}{\Lambda^2} \bar{L}_L \varphi
\,\sigma^{\mu\nu} \,\tau_R B_{\mu\nu} + \frac{C_{\tau W}}{\Lambda^2}
\bar{L}_L \varphi \,\vec{\tau} \,\sigma^{\mu\nu}\tau_R
\vec{W}_{\mu\nu} + \,\mathrm{h.c.} \;,
\end{equation}
$L_L = (\nu_L, \tau_L)$ is the tau leptonic doublet, $\varphi$ is
the Higgs doublet, $B_{\mu\nu}$ and $W_{\mu\nu}$ are the $U(1)_Y$
and $SU(2)_L$ field strength tensors. One can also write the
dimension-6 operator
\begin{equation}\label{operator}
\frac{C_{\tau B}}{\Lambda^2} \bar{L}_L \sigma^{\mu\nu} \hat{D} L_L
B_{\mu\nu} \;,
\end{equation}
with $\hat{D} = \gamma^\mu D_\mu$. However, this operator reduces to
the first operator in \eqref{L_SMEFT} after using the equations of
motion.

After electroweak symmetry breaking, BSM contributions to the tau
magnetic and electric dipole moments can be written as follows
\begin{equation}\label{MMs_mod}
\Delta a_\tau = \frac{2m}{e} \frac{\sqrt{2}\,v}{\Lambda^2}
\mathfrak{Re}(C_{\tau\gamma}) \;, \quad d_\tau =
\frac{\sqrt{2}\,v}{\Lambda^2} \mathfrak{Im}(C_{\tau\gamma}) \;,
\end{equation}
where $v = 246$ GeV,
\begin{equation}\label{C_gamma}
C_{\tau\gamma} = \cos\theta \,C_{\tau B} - \sin\theta \,C_{\tau W}
\;,
\end{equation}
and $\theta$ is the Weinberg angle. Thus, the complex couplings
break CP conservation and lead to the electric dipole moment.

In collider based measurements of tau lepton, the photon is
off-shell.  However, when the BSM scale $\Lambda^2$ is much higher
than $q^2$, terms of higher orders of $q^2/\Lambda^2$ can be
neglected in the form factor expansion, and the collider
measurements can be interpreted through eq.~\eqref{zero_q2_MM} into
constraints on $\delta a_\tau = a_\tau$, $\delta d_\tau = d_\tau$.

Note that interactions \eqref{L_SMEFT} necessarily imply a
modification of the coupling of the Higgs field $h$ to $\tau$,
\begin{equation}\label{tau_Higgs}
\frac{C_{\tau h}}{\Lambda^2} h \bar{\tau} \sigma^{\mu\nu} \tau
F_{\mu\nu} \;,
\end{equation}
where $C_{\tau h}$ is a linear combination of the couplings $C_{\tau
W}$  and $C_{\tau B}$. Thus, rare Higgs decays offer a probe of
$\tau$ anomalous magnetic moment as well \cite{Howard:2019}.

The SMEFT dipole operators induce a momentum-dependent modification
of the $\gamma^*\tau\tau$ vertex, effectively probing the form
factors $F_2(q^2)$ and $ F_3(q^2)$ away from the static limit ($q^2
\rightarrow 0$). As a result, the extracted limits do not correspond
to a direct measurement of the tau magnetic moments, but rather to
bounds on the coefficients $C_{\tau B}$, $C_{\tau W}$ in
\eqref{L_SMEFT}.




\end{document}